# Piezoelectric Response and Free Energy Instability in the Perovskite Crystals BaTiO$_3$, PbTiO$_3$ and Pb(Zr,Ti)O$_3$

Marko Budimir, Dragan Damjanovic[*], and Nava Setter

*Ceramics Laboratory, Materials Science and Engineering Institute,*

*Ecole polytechnique fédérale de Lausanne - EPFL, 1015 Lausanne, Switzerland*

**Abstract:** The question of the origin of the piezoelectric properties enhancement in perovskite ferroelectrics is approached by analyzing the Gibbs free energy of tetragonal BaTiO$_3$, PbTiO$_3$ and Pb(Zr,Ti)O$_3$ in the framework of the Landau-Ginzburg-Devonshire theory. The flattening of the Gibbs free energy profile appears as a fundamental thermodynamic process behind the piezoelectric enhancement. The generality of the approach is demonstrated by examining the free energy flattening and piezoelectric enhancement as a function of composition, temperature, electric field and mechanical stress. It is shown that the anisotropy of the free energy flattening is the origin of the anisotropic enhancement of the piezoelectric response, which can occur either by polarization rotation or by polarization contraction. Giant enhancement of the longitudinal piezoelectric response ($\approx 10^3$ pC/N) is predicted in PbTiO$_3$ under uniaxial compression.

PACS: 77.84.Dy, 77.65.Bn

---

[*] electronic address: dragan.damjanovic@epfl.ch

**I. Introduction**

The piezoelectric properties of perovskite crystals have been of considerable interest in the last several years. Unexpectedly large piezoelectric response along nonpolar crystallographic directions was reported in complex ferroelectrics by Kuwata *et al.*[1], Park and Shrout[2], and Du *et al.*[3] The discovery of the effect in simple perovskites[4, 5] has suggested that this behavior is a common characteristic of perovskite ferroelectrics. The origin of the enhancement, whose technological impact is potentially significant, has been traced to polarization rotation under external electric field.[6] Subsequent studies have demonstrated that the enhancement is dependent on the density of domain walls in crystals with an engineered domain structure,[7, 8] hierarchical domain structure,[9] electric field[10] and stress[11] induced phase transitions, proximity of the phase transition temperatures,[12, 13] crystal instability induced by electric field applied antiparallel to polarization,[14] compressive stress applied along the polar axis,[15, 16] and composition/morphotropic phase boundary effects.[17-20] Considering importance of this effect from both the fundamental and technological point of view, it is clearly of interest to see if there exists a common underlying process for piezoelectric enhancement in ferroelectric perovskites.

In this paper, the problem is addressed in the framework of the Landau-Ginzburg-Devonshire (LGD) theory. It is shown that on the thermodynamic phenomenological level the common origin of the piezoelectric enhancement and its anisotropy in perovskites is the flattening of a free energy profile. The relationship between the enhancement of the piezoelectric response and a free energy profile flattening was introduced by Fu and Cohen using a first principles approach.[6] Recently, Wu and Cohen[11] made a link between the enthalpy difference between two pressure-induced crystal phases and enhanced piezoelectric response in PbTiO$_3$. In both cases considered by Cohen and co-workers the large piezoelectric response is related to the field (electrical or mechanical) induced phase transitions and polarization rotation.

Here we demonstrate universal applicability of this idea. We discuss effects of composition, stress, electric field and temperature on the Gibbs free energy instability in Pb(Zr,Ti)O$_3$, BaTiO$_3$ and PbTiO$_3$ and ensuing enhancement of the piezoelectric response. In contrast to work of Cohen and co-workers, we show that phase transitions related free energy instabilities are not the only way to achieve large enhancement of the piezoelectric response. Huge enhancement of the piezoelectricity can be expected in the vicinity of and just below the thermodynamic coercive fields within the same ferroelectric phase. In the case of PbTiO$_3$ under uniaxial compressive stress, we predict huge longitudinal piezoelectric coefficient comparable to that in relaxor ferroelectrics. It is also shown that the anisotropy of the free energy profile determines whether the enhancement of piezoelectricity will take place by polarization rotation or polarization contraction. We also show that, in contrast to rhombohedral perosvkites, in the case of tetragonal Pb(Zr,Ti)O$_3$ materials lying in the vicinity of the morphotropic phase boundary, the composition related flattening of the free energy profile and piezoelectric enhancement are isotropic i.e. effects of polarization rotation and polarization contraction are comparable. This isotropy can be broken by external stresses and electric fields leading to large, polarization rotation related enhancement of the piezoelectric effect.

**II. Results and discussion**

For the purposes of this paper, we investigate the Gibbs free energy and the longitudinal piezoelectric coefficient of the tetragonal phase of perovskite BaTiO$_3$ and (1-x)PbTiO$_3$-

xPbZrO$_3$ (PZT) monodomain single crystals. In both materials the ferroelectric tetragonal phase exhibits *4mm* and the paraelectric cubic phase *m3m* symmetry; the polar axis is oriented along the [001] direction of the cubic system. In the framework of the LGD theory,[21, 22] the Gibbs free energy $\Delta G$ can be written as the series expansion of the polarization $P = (P_1, P_2, P_3)$. While all calculations in this article are concerned with the tetragonal phase ($P_1 = P_2 = 0; P_3 \neq 0$), the proximity of the orthorhombic phase ($P_1 = 0; P_2 = P_3 \neq 0$) is taken into account in BaTiO$_3$ near the tetragonal-orthorhombic phase transition temperature. Similarly, as shown elsewhere,[16] variation of the $\Delta G$ with $P_2$ for equilibrium $P_3$ gives susceptibility of a tetragonal ferroelectric to polarization rotation and tendency toward a monoclinic distortion. In a more general case all three components of polarization may be included in the analysis,[23] however, such generalization is beyond the scope of the present paper.

If external electric and elastic fields are applied along the $P_3$, the Gibbs free energy can be written in the coordinate system of the cubic phase as:[21]

$$\Delta G = \alpha_1(P_2^2 + P_3^2) + \alpha_{11}(P_2^4 + P_3^4) + \alpha_{12}P_2^2P_3^2 + \alpha_{111}(P_2^6 + P_3^6) + \alpha_{112}(P_2^4P_3^2 + P_3^4P_2^2) \\ - s_{11}^D\sigma_3^2/2 - Q_{11}\sigma_3P_3^2 - Q_{12}\sigma_3P_2^2 - E_3P_3 \quad (1)$$

where $\alpha$s are the dielectric stiffness coefficients, $\sigma_3$ and $E_3$ are respectively the stress and the electric field components along the polar axis, $s_{11}^D$ is the elastic compliance at constant polarization, and $Q_{ij}$ are electrostrictive constants. The values of the $\alpha$ and $Q_{ij}$ coefficients are taken from Refs.[21, 24, 25] At all examined temperatures $s_{11}^D$ is taken as $9 \times 10^{-12}$ m$^2$/N for BaTiO$_3$[26] and as $6.785 \times 10^{-12}$ m$^2$/N for Pb(Zr,Ti)O$_3$.[27] The negative $\sigma_3$ and $E_3$ have the meaning of compressive stress and electric field applied antiparallel to polarization.[28]

The dielectric susceptibility is calculated as $\chi_{ij} = \left[\partial^2 \Delta G / \partial P_i \partial P_j\right]^{-1}$ and the longitudinal, the transverse and the shear piezoelectric coefficients as, respectively: $d_{33} = 2\varepsilon_0 \chi_{33} Q_{11} P_3$, $d_{31} = 2\varepsilon_0 \chi_{33} Q_{12} P_3$ and $d_{15} = \varepsilon_0 \chi_{11} Q_{44} P_3$, where $\varepsilon_0$ is the permittivity of vacuum.[21] $P_3$ and $\chi_{ij}$ are functions of $\sigma_3$, $E_3$[29] and temperature. The origin of the temperature dependence is in Curie-Weiss behavior of $1/\alpha_1$; in BaTiO$_3$ higher order dielectric stiffnesses are also temperature dependent.[25, 30] It is easily seen that the flattening of a simple polynomial such as $\Delta G$ in Eq. (1) implies flattening and decrease of its first and second derivatives. Since $d \propto \chi$ and $\chi_{ij} = \left[\partial^2 \Delta G / \partial P_i \partial P_j\right]^{-1}$, the flattening of the free energy profile implies increase of the system's dielectric susceptibility, and thus the increase of its piezoelectric response. To calculate effects away from the crystallographic axes, the orientation dependence of the longitudinal piezoelectric coefficient, $d_{33}^*$, of a tetragonal crystal may be expressed as: $d_{33}^*(\theta) = d_{33} \cos^3 \theta + (d_{31} + d_{15}) \cos\theta \sin^2\theta$, where $\theta$ is the angle between the new arbitrary direction and the polar axis.[13] Thus, $d_{33}^*(\theta)$ is a function of both the susceptibility along the polar axis, $\chi_{33}$, and the susceptibility perpendicular to it, $\chi_{11}$. This dependence is the basis for $d_{33}^*(\theta)$ enhancement driven by either polarization rotation ($\propto \chi_{11}$)[11, 14, 16, 20] or the polarization contraction ($\propto \chi_{33}$).[13, 14, 16]

To illustrate relationship between the Gibbs free energy flattening and the enhancement of the $d_{33}^*$, we analyze $\Delta G$ and $d_{33}^*$ of PZT as a function of the Zr/Ti ratio, compressive stress and electric field applied along the polar axis. In BaTiO$_3$, $\Delta G$ and $d_{33}^*$ are analyzed as functions of temperature and electric field applied along the polar axis.

We first consider effects of Zr/Ti ratio on the Gibbs free energy and piezoelectric response in PZT at 298 K. Two tetragonal compositions, one with Zr/Ti=0/100 (i.e., pure PbTiO$_3$) and the other with Zr/Ti=40/60 are considered. These two compositions are chosen to illustrate effects of the proximity of the morphotropic phase boundary (MPB) on the Gibbs free energy profile flattening and the piezoelectric enhancement. In PZT, the MPB appears at Zr/Ti=52/48 and PZT 40/60 is sufficiently far from it[31] that the complications arising from possible presence of mixed phases or a monoclinic phase can be avoided. Two different free energy profiles were calculated: one along the polar direction with $P_2 = 0; P_3 \neq 0$ and the other with $P_2 \neq 0; P_3 = P_3(\sigma_3, E_3)$, were $P_3$ is fixed at its equilibrium value at 298 K. The former case (Fig. 1a) involves elongation ($E_3 > 0$) or contraction ($E_3 < 0$) of the polarization along the polar axis, while the latter case (Fig. 1b) corresponds to the polarization rotation away from the polar direction, as described in detail in Ref. 23. As expected,[32] both profiles are flatter in PZT 40/60 lying closer to the MPB. As a consequence, $\chi_{33}$ and $\chi_{11}$ are larger in PZT 40/60 than in PZT 0/100 leading to enhancement of the corresponding piezoelectric coefficients, Fig. 1c. This enhancement of the properties in compositions close to the MPB is a well-known empirical and theoretical result,[24] interpreted here in terms of the Gibbs free energy flattening. Significantly, the analysis shows that the flattening of the $\Delta G$ profile away from the polar axis and along the polar axis are comparable. In fact, the $d_{33}^*(\theta)$ surface is elongated along the polar axis, Figs. 1c, indicating that the maximum enhancement is along the polar axis ($\theta = 0°$). This is qualitatively different from the behavior of the rhombohedral phases of PZT,[3, 20] BaTiO$_3$,[6] and relaxor ferroelectrics,[2] where piezoelectric enhancement is strongest along nonpolar directions. We next show that under external electric field and stress applied against polarization, the isotropy of the free energy profile is broken, leading to a large enhancement of the piezoelectricity by polarization rotation.

The effect of the electric field bias $E_3$ and the mechanical stress $\sigma_3$ on the $\Delta G$ and $d_{33}^*(\theta)$ in PZT 40/60 composition is shown in Fig. 2. Figures 2a-2c compare the $\Delta G$ and $d_{33}^*(\theta)$ for the crystal at zero bias field and for $E_3$= -35, -43 and -44 MV/m applied antiparellel to the polarization. Likewise, Figs. 2d-2f show the $\Delta G$ and $d_{33}^*(\theta)$ for uncompressed crystal and for the crystal subjected to compressive stress of $\sigma_3$ = -350 and -500 MPa applied along the polar direction. In the limits of the phenomenological theory used, neither the electric field nor the stress are high enough to cause polarization switching by 180° or 90°; thus, the crystals remain in the tetragonal single domain state.[14, 16]

At low electric fields and compressive pressures the dominant enhancement of the $d_{33}^*(\theta)$ is along the polar direction, i.e. it is a consequence of the colinear polarization contraction. This behavior changes dramatically at high antiparellel electric fields and compressive pressures approaching thermodynamic coercive fields, where instability of the $\Delta G$ and $d_{33}^*(\theta)$ enhancement become strongly anisotropic and polarization rotation effects dominate the piezoelectric response. At high fields (compare $\Delta G$ and $d_{33}^*(\theta)$ for $E_3$= -43 and -44 MV/m in Fig. 2a-c) even a small increase in the flatness of the $\Delta G$ profile leads to a large enhancement of $d_{33}^*(\theta)$ along off-polar directions.

Remarkably, our calculations show that PbTiO$_3$ exhibits giant enhancement of the $d_{33}^*(\theta)$ along nonpolar directions once compressive stress is sufficiently large, Fig. 3. At 300 K and $\sigma_3$ = -1.79 GPa, just below the thermodynamic coercive pressure, the value of the noncolinear $d_{33}^*(\theta)$ in PbTiO$_3$ is on the order of $10^3$-$10^4$ pC/N. This result is qualitatively similar to that obtained by Wu and Cohen using *ab initio* calculations for PbTiO$_3$ under hydrostatic pressure.[11] We emphasize, however, that in the two cases the origin of the free energy instability is different. In the report by Wu and Cohen, the instability is related to the presence of the morphotropic phase boundary that is induced by the hydrostatic pressure. At this boundary the polarization changes direction from pseudocubic [001] axis in the tetragonal phase to pseudocubic [111] axis in the rhombohedral phase. In our work, the instability is related to the multiple orientations of the ferroelectric polarization within the same crystal phase (i.e., switching by 90° in tetragonal materials from [001] to [010] axis). Once the thermodynamic coercive compressive pressure is approached *but not passed,* the crystal is destabilized, the free energy becomes shallow and the piezoelectricity is greatly enhanced.

Interestingly, the polarization rotation effects under antiparallel electric fields in PbTiO$_3$ are small. The reasons for this difference between effects of compressive pressure and antiparallel field on piezoelectric anisotropy will be discussed elsewhere.

Recent experimental studies of effects of hydrostatic pressure on 52/48 PZT MPB composition[33, 34] and studies of effects of tensile stress perpendicular to polarization in thin films of PbTiO$_3$[35] indicate that polarization rotation indeed occurs in these material and that the symmetry becomes lower in both compositions under the effect of stress. We cannot predict monoclinic or triclinic phases in the framework of the 6$^{th}$ order LGD theory;[36] however, as indicated above and explained in more details in Ref. 23, our approach does indicate susceptibility of materials to monoclinic distortion, correctly predicting the relationship between the instability of the $\Delta G$, susceptibility of the material to polarization rotation and ensuing enhancement of the $d_{33}^*(\theta)$.

We next show that the proposed approach is also applicable to processes involving temperature driven enhancement of the piezoelectric response. As an example, we analyze effects of electric field and temperature on the $\Delta G$ and $d_{33}^*(\theta)$ in the tetragonal phase of BaTiO$_3$, in a vicinity of 298 K. Figures 4a-b illustrate the flattening of the $\Delta G$ profile by application of a bias electric field antiparallel to polarization at 298 K. The behavior is similar to that predicted in BaTiO$_3$ at high compressive pressures[16] and in PZT 40/60 and PbTiO$_3$ shown above: the flattening of the $\Delta G$ profile is anisotropic, being stronger away from the polar axis, Fig. 4b, than along the polar axis, Fig. 4a. This leads to the maximum $d_{33}^*(\theta)$ approximately along the [111] axis, Fig. 4c, while its enhancement along the polar axis [001] is comparatively smaller. Finally, it is interesting that the free energy anisotropy is strongly influenced by the proximity of the tetragonal-orthorhombic phase transition temperature that occurs at 283 K. At this phase transition temperature the polarization vector changes its direction from [001] ($P_2 = 0, P_3 \neq 0$) to [011] ($P_2 = P_3 \neq 0$) leading to the dielectric softening of the crystal in the plane perpendicular to the polar axis and to an increase in $\chi_{11}$[13]. The dominant temperature effect is clearly seen in Figs. 4d-f which show that the anisotropic flattening of the $\Delta G$ profile and enhancement of the piezoelectric response at zero field become stronger as the tetragonal-orthorhombic phase transition temperature is approached on cooling. The antiparallel electric field in Figs. 4a-c has therefore an additional destabilizing effect on the crystal and enhances the temperature driven anisotropic flattening of the $\Delta G$ profile and enhancement of the $d_{33}^*(\theta)$ along a nonpolar axis.

**III. Conclusions**

This work emphasizes an important fundamental fact that is common for several types of piezoelectric properties enhancement reported in perovskite crystals. It is shown that the flattening of the Gibbs free energy profile, regardless of whether it is caused by temperature or composition variation, application of compressive pressure or antiparellel electric field bias, leads to enhancement of the dielectric susceptibilities and the piezoelectric response of examined materials. The universality of the approach is further indicated by the fact that the anisotropic flattening of the free energy profile can explain both enhancement of the piezoelectric response by polarization rotation away from the polar axis and by polarization contraction along the polar axis. The relatively isotropic free energy profile in tetragonal PZT, even in compositions close to the MPB, can be broken by external electric fields and compressive stresses applied antiparallel to polarization. The huge enhancement of the piezoelectric response by compressive stress predicted in PbTiO$_3$ along off-polar directions suggests that compositional or structural disorder, such as present in complex relaxor-ferroelectrics, is not essential for the giant piezoelectric effect. Our results and those of *ab initio* calculations[11] hint that such disorder is probably responsible for the free energy instability, which, as shown here, leads to a large piezoelectric response.

The authors acknowledge financial support of the Swiss National Science Foundation.

References:

[1] J. Kuwata, K. Uchino, and S. Nomura, Jpn. J. Appl. Phys. 21, 1298 (1982).

[2] S. E. Park and T. R. Shrout, J. Appl. Phys. 82, 1804 (1997).

[3] X. Du, U. Belegundu, and K. Uchino, Jpn. J. Appl. Phys. 36, 5580 (1997).

[4] S. Wada, S. Suzuki, T. Noma, T. Suzuki, M. Osada, M. Kakihana, S.-E. Park, L. E. Cross, and T. R. Shrout, Jpn. J. Appl. Phys. 38, 5505 (1999).

[5] K. Nakamura, T. Tokiwa, and Y. Kawamura, J. Appl. Phys. 91, 9272 (2002).

[6] H. Fu and R. E. Cohen, Nature 403, 281 (2000).

[7] S. Wada, K. Yako, H. Kakemoto, T. Tsurumi, and T. Kiguchi, Journal of Applied Physics 98, 014109 (2005).

[8] R. Ahluwalia, T. Lookman, A. Saxena, and W. W. Cao, Physical Review B 72, 014112 (2005).

[9] Y. M. Jin, Y. U. Wang, A. G. Khachaturyan, J. F. Li, and D. Viehland, Physical Review Letters 91, 197601 (2003).

[10] L. Bellaiche, A. Garcia, and D. Vanderbilt, Phys. Rev. B 64, 060103 (2001).

[11] Z. G. Wu and R. E. Cohen, Physical Review Letters 95, 037601 (2005).

[12] Y. Lu, D.-Y. Jeong, Z.-Y. Cheng, Q. M. Zhang, H.-S. Luo, Z.-W. Yin, and D. Viehland, Appl. Phys. Lett. 78, 3109 (2001).

[13] M. Budimir, D. Damjanovic, and N. Setter, J. Appl.Phys. 94, 6753 (2003).

[14] M. Budimir, D. Damjanovic, and N. Setter, Appl. Phys. Lett. 85, 2890 (2004).

[15] D. Viehland, J. F. Li, E. McLaughlin, J. Powers, R. Janus, and H. Robinson, Journal of Applied Physics 95, 1969 (2004).

[16] M. Budimir, D. Damjanovic, and N. Setter, Phys. Rev. B 72, 064107 (2005).

[17] Y. Ishibashi and M. Iwata, Jpn. J. Appl. Phys. 37, L985 (1998).

[18] M. Iwata, H. Orihara, and Y. Ishibashi, Ferroelectrics 266, 57 (2002).

[19] R. Guo, L. E. Cross, S.-E. Park, B. Noheda, D. E. Cox, and G. Shirane, Phys. Rev. Lett. 84, 5423 (2000).

[20] L. Bellaiche, A. Garcia, and D. Vanderbilt, Phys. Rev. Lett. 84, 5427 (2000).

[21] M. J. Haun, E. Furman, S. J. Jang, H. A. McKinstry, and L. E. Cross, J. Appl. Phys. 62, 3331 (1987).

[22] A. F. Devonshire, Phil. Mag. 40, 1040 (1949).

[23] D. Damjanovic, J. Am. Ceram. Soc. 88, 2663 (2005).

[24] M. J. Haun, E. Furman, S. J. Jang, and L. E. Cross, Ferroelectrics 99, 63 (1989).

[25] A. J. Bell and L. E. Cross, Ferroelectrics 59, 197 (1984).

[26] A. Schaefer, H. Schmitt, and A. Dörr, Ferroelectrics 69, 253 (1986).

[27] A. Amin, R. E. Newnham, and L. E. Cross, Phys. Rev. B 34, 1595 (1986).

[28] A. Amin, L. E. Cross, and R. E. Newnham, Ferroelectrics 99, 145 (1989).

[29] L. Chen, V. Nagarajan, R. Ramesh, and A. L. Roytburd, Journal of Applied Physics 94, 5147 (2003).

[30] J. Iniguez, S. Ivantchev, J. M. Perez-Mato, and A. Garcia, Phys. Rev. B 63, 144103 (2001).

[31] B. Noheda, D. E. Cox, G. Shirane, J. A. Gonzalo, L. E. Cross, and S.-E. Park, Appl. Phys. Lett. 74, 2059 (1999).

[32] J.-C. Toledano and P. Toledano, The Landau Theory of Phase Transitions (World Scientific, Singapore, 1987).

[33] J. Rouquette, J. Haines, V. Bornand, M. Pintard, P. Papet, C. Bousquet, L. Konczewicz, F. A. Gorelli, and S. Hull, Phys. Rev. B 70 (2004).

[34] J. Rouquette, J. Haines, V. Bornand, M. Pintard, P. Papet, W. G. Marshall, and S. Hull, Physical Review B 71, 024112 (2005).

[35] G. Catalan, A. Janssens, G. Rispens, S. Csiszar, O. Seeck, G. Rijnders, D. H. A. Blank, and B. Noheda, Phys. Rev. Lett. 96, 127602 (2006).

[36] D. Vanderbilt and M. H. Cohen, Phys. Rev. B 63, 094108 (2001).

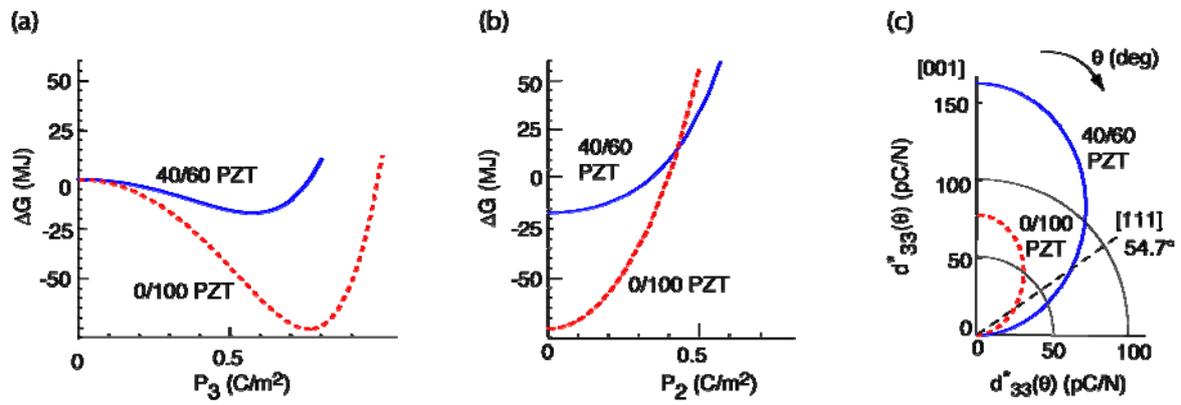

Figure 1. (color online) Effects of composition in PZT 0/100 (PbTiO$_3$) and PZT 40/60 at 298 K on free energy flattening and piezoelectric response: (a) $\Delta G(P_2 = 0, P_3)$ profile related to polarization contraction; (b) $\Delta G(P_2, P_3 = 0.52$ C/m$^2)$ profile indicating polarization rotation; (c) polar plot of $d_{33}^*(\theta)$.

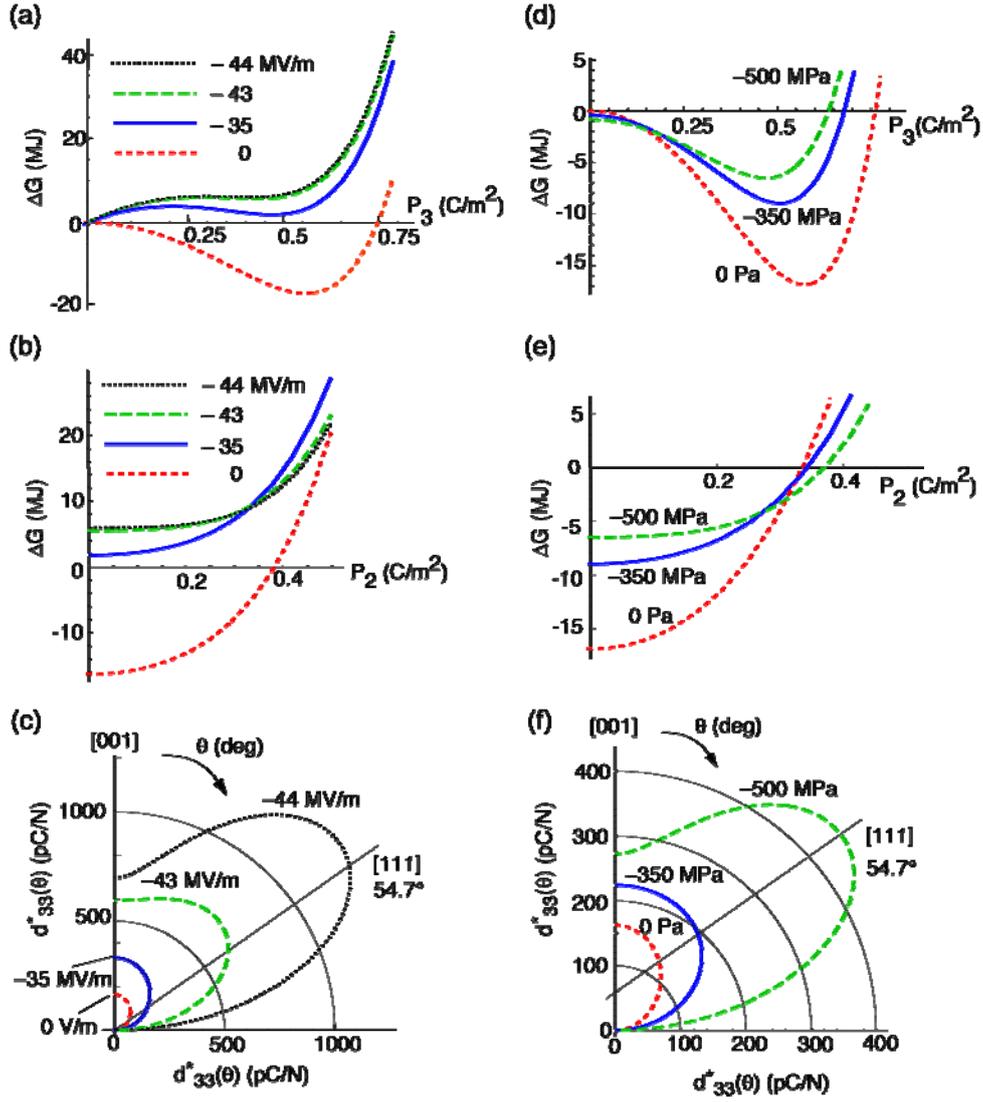

Figure 2. (color online) Effects of electric bias field ($E_3$= 0, -35, -43 and -44 MV/m) and compressive stress ($\sigma_3$= 0, -350, and -500 MPa) on free energy profile and $d_{33}^*(\theta)$ in PZT 40/60 at 298 K: a) and d) $\Delta G(P_2 = 0, P_3)$ profile indicating polarization contraction; (b) $\Delta G[P_2, P_3 = \text{const}(E_3)]$ and (e) $\Delta G[P_2, P_3 = \text{const}(\sigma_3)]$ profiles indicating polarization rotation; (c) and (f) are polar plots of $d_{33}^*(\theta)$. In (a) and (b) only $\Delta G(P_2 = 0, P_3 \geq 0)$ is of interest and is shown. (c)-(f) are symmetrical with respect to the vertical axis. Coercive electric field for this composition is just above -44 MV/m. In highly unstable regions even small changes in the flatness of the $\Delta G$ (compare $\Delta G$ and $d_{33}^*(\theta)$ for $E_3$= -43 and -44 MV/m) can have a huge impact on $d_{33}^*(\theta)$.

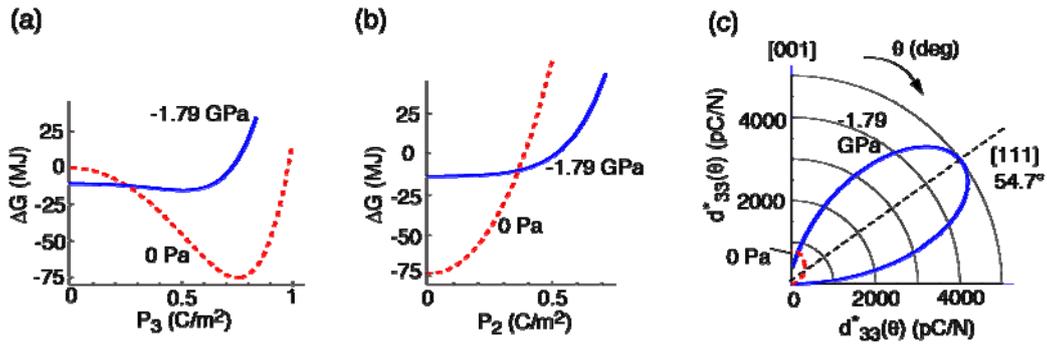

Figure 3. (color online) Effect of compressive stress ($\sigma_3 = 0$ and -1.79 GPa) at T=300 K on anisotropic free energy flattening and piezoelectric enhancement in PbTiO$_3$: (a) $\Delta G(P_2 = 0, P_3)$ profile indicating polarization contraction; (b) $\Delta G[P_2, P_3 = \text{const.}(\sigma_3)]$ profile indicating polarization rotation; (c) polar plots of $d^*_{33}(\theta)$. At $\sigma_3 = 0$ Pa values for $d^*_{33}(\theta)$ (dashed line) is multiplied by 10. For $\sigma_3$ close to the coercive stress (approx. -1.9 GPa) values of $d^*_{33}(\theta)$ are strongly sensitive to input parameters and vary between $10^3$ and $10^4$.

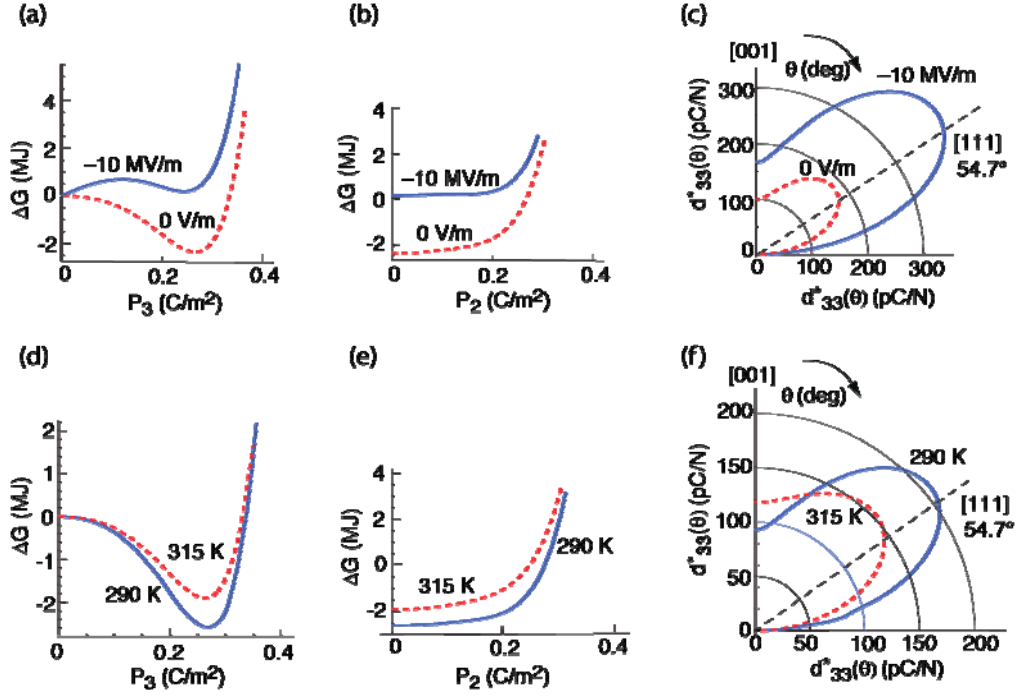

Figure 4. (color online) Effects of (a-c) electric bias field ($E_3$=0, -10 MV/m) at T= 298 K and (d-f) proximity of tetragonal-orthorhombic phase transition temperature (T=283 K) at zero field on anisotropic free energy flattening and piezoelectric enhancement in BaTiO$_3$: (a) and (d) $\Delta G(P_2 = 0, P_3)$ profile indicating polarization contraction; (b) and (e) $\Delta G[P_2, P_3 = \text{const.}(E_3, T)]$ profile indicating polarization rotation; (c) and (f) are polar plots of $d_{33}^*(\theta)$. Note that the electric field reinforces temperature driven effects. Compressive stress effects in BaTiO$_3$ are reported in Ref. 16.